\documentclass[twocolumn,showpacs,aps]{revtex4}
\usepackage{graphicx}

\begin{document}
\title{Complementary contributions of indeterminism and signalling to quantum correlations}

\author{Michael J.W. Hall}
\affiliation{Theoretical Physics, Research School of Physics and Engineering,  Australian National
University, Canberra ACT 0200, Australia}


\begin{abstract}
Simple quantitative measures of indeterminism and signalling, $I$ and $S$, are defined for models of statistical correlations.  It is shown that any such model satisfies a generalised Bell-type inequality, with tight upper bound $B(I,S)$.  This upper bound explicitly quantifies the complementary contributions required from indeterminism and signalling, for modelling any given violation of the standard Bell-CHSH inequality. For example, all models of the maximum quantum violation must either assign no more than 80\% probability of occurrence to some underlying event, and/or allow a nonlocal change of at least 60\% in an underlying marginal probability of one observer in response to a change in measurement setting by a distant observer.  The results yield a corresponding complementarity relation between the numbers of local random bits and nonlocal signalling bits required to model a given  violation. A stronger relation is conjectured for simulations of singlet states. Signalling appears to be a useful resource only if a `gap' condition is satisfied, corresponding to being able to nonlocally flip some underlying marginal probability $p$ to its complementary value $1-p$.
\end{abstract}

\maketitle

\section{Introduction} 

Bell inequalities are of great physical interest, because they demonstrate that certain sets of very plausible physical properties are incompatible with quantum correlations \cite{bell}.  In particular, any violation of a Bell inequality requires that at least one of these plausible properties must be relaxed.  The question then naturally arises: by how much?
This question is investigated here in the context of relaxing no-signalling and determinism properties.  

A correlation model is {\it no-signalling} if all underlying outcome probabilities for one detector are independent of the measurement setting of any distant detector.  Without this property, communication between distant observers is possible, in principle, via one observer varying a local measurement setting. It holds, for example, in the standard Hilbert space model of quantum correlations \cite{hg}, but not in the deBroglie-Bohm model \cite{dbb}, nor in the Toner-Bacon model of the singlet state \cite{bt}.

Further, a correlation model is {\it deterministic} if the observed statistical correlations are generated by averages over a set of underlying or `hidden' variables (traditionally denoted by $\lambda$), where for any fixed underlying variable all measurement outcomes are fully determined.  It is natural to postulate this property in the case of perfect correlations between outcomes \cite{bell,epr,ghz}.  It holds, for example, in the deBroglie-Bohm and Toner-Bacon models of quantum correlations, but not in the standard Hilbert space model.

Various types of Bell inequalities may be derived for any correlation model satisfying the above properties of no-signalling and determinism, providing it is assumed that measurement settings may be chosen independently of the underlying variables \cite{bell, chsh, bell1}.  This assumption corresponds to the notion that there is no `universal conspiracy' preventing observers from freely choosing measurement settings \cite{shimony}, and will be retained in what follows (except in the final paragraph).  The main question to be addressed here is then {\it by how much do the properties of no-signalling and determinism need to be relaxed, to be able to model quantum correlations}?  

To obtain quantitative results, it is of course necessary to define measures of the degree of indeterminism, $I$, and degree of signalling, $S$, for a given underlying model of statistical correlations.   Here, $I$ will be chosen as the smallest positive number for which every underlying outcome probability, $p$, satisfies $p\in [0,I]\cup[1-I,1]$. Thus, $I\leq 1/2$, and $I=0$ if and only if the underlying probability distributions are deterministic, with $p$ confined to $\{0,1\}$.
Further, $S$ will be chosen as the maximum change in an underlying outcome probability for one detector, in response to a change of measurement setting of another detector.  Thus, $0\leq S\leq 1$, with $S=0$ if and only if no signalling is possible between the detectors via the choice of measurement settings.  

The above definitions are advantageous in that they have direct interpretations in terms of probabilities, and their respective contributions to `quantum nonlocality' can be calculated in a straightforward manner. It will be seen that these contributions are {\it complementary} in nature.   For example, it is known that any no-signalling model of the singlet state must be maximally indeterministic \cite{branc}, i.e., that $S=0$ implies $I=1/2$ for such models.  Here, solid support and a partial proof will be given for the more general conjecture that
\begin{equation} \label{comp}
\ S+2I\geq 1 
\end{equation}
for {\it any} model of the singlet state. The conjecture may be also be formulated as a complementarity relation between the numbers of local random bits and nonlocal signalling bits required to model a singlet state.

Further, a theorem is derived, in the form of a `relaxed' Bell inequality, that bounds the statistical correlations of models having nonzero values of $I$ and $S$.   This places strong constraints on the minimum possible degrees of indeterminism and signalling that must be present in any model that violates a standard Bell inequality.  These constraints may, similarly to Eq.~(\ref{comp}), be regarded as complementarity relations, and can also be alternatively formulated as trade-off relations between the numbers of local random bits and nonlocal signalling bits required to model a given Bell inequality violation.

In the following section, the above degrees of indeterminism and signalling, $I$ and $S$, are rigorously defined for any underlying model of correlations.  Sec.~III presents the relaxed Bell inequality and consequences thereof, with a proof deferred to the Appendix; Sec.~IV obtains information-theoretic forms of the complementarity relations; Sec.~V provides support for the singlet state conjecture in Eq.~(\ref{comp}); and some general discussion is given in Sec.~VI.
 
\section{Quantifying indeterminism and signalling}
 
Consider a set of statistical correlations defined by joint probabilities,  $p_{XY}(a,b)$, corresponding to outcomes $a$ and $b$ for respective measurement settings $X$ and $Y$.  Any model of the correlations introduces underlying variables, $\lambda$, and underlying joint probabilities $p_{XY}(a,b|\lambda)$ conditional on $\lambda$.   Bayes theorem implies that 
\begin{equation} \label{underlying}
p_{XY}(a,b) = \int d\lambda \, \rho_{XY}(\lambda)\, p_{XY}(a,b|\lambda) ,
\end{equation}
where $\rho_{XY}(\lambda)$ is a probability density over the underlying variables.  
As noted in the introduction, it will be assumed that the measurement settings are independent of the underlying variables, implying that 
\[ \rho_{XY}(\lambda)\equiv \rho(\lambda). \]
For given $X$, $Y$ and $\lambda$, the marginal probability distributions for the first and second detectors will be denoted by $p^{(1)}_{XY}(a|\lambda):=\sum_b p_{XY}(a,b|\lambda)$ and $p^{(2)}_{XY}(b|\lambda):=\sum_a p_{XY}(a,b|\lambda)$, respectively.  

No-signalling is the property that the underlying marginal probabilities for one detector do not depend on the measurement setting of the other detector:
\begin{equation} \label{nosig}
 p^{(1)}_{XY}(a|\lambda) = p^{(1)}_{XY'}(a|\lambda) ,~~
 p^{(2)}_{XY}(b|\lambda) = p^{(2)}_{X'Y}(b|\lambda) .
\end{equation}
Thus, even with full knowledge of $\lambda$, no information can in principle be transmitted between observers via a choice of measurement settings.  This property, also known as parameter independence, is consistent with the principle of relativistic locality in the case that the observers perform their measurements in spacelike separated regions.  

Determinism is the property that all underlying probabilities are confined to the values 0 and 1.
Thus, given full knowledge of $\lambda$, all measurement outcomes are known with certainty.  Since a joint distribution is deterministic if and only if the marginals are deterministic, this property may be stated formally as 
\begin{equation} \label{det}
p^{(1)}_{XY}(a|\lambda), ~~p^{(2)}_{XY}(b|\lambda) \in \{0,1\} .
\end{equation}

Natural {\it one-way} degrees of signalling are now defined by the maximum extent to which one observer can influence an outcome probability for the other observer, via a choice of measurement settings, i.e.,
\begin{eqnarray} \label{1to2}
S_{1\rightarrow 2} &:=& \sup_{\{X,X',Y,b,\lambda\} } \left| p^{(2)}_{XY}(b|\lambda) - p^{(2)}_{X'Y}(b|\lambda) \right|,\\ \label{2to1}
S_{2\rightarrow 1} &:=& \sup_{\{X,Y,Y',a,\lambda\} } \left| p^{(1)}_{XY}(a|\lambda) - p^{(1)}_{XY'}(a|\lambda) \right| .
\end{eqnarray}
Clearly, these vanish if and and only the no-signalling property holds.
Further, a natural {\it local} degree of indeterminism, $I_j$, is given by the smallest positive number such that the corresponding marginal probabilities lie in $[0,I_j]\cup [1-I_j,1]$, i.e., 
\begin{equation} \label{ilocal}
I_j := \sup_{\{X,Y,\lambda\}}     \min_z \left\{p^{(j)}_{XY}(z|\lambda),1-p^{(j)}_{XY}(z|\lambda) \right\} ,
\end{equation}
for $j=1,2$. Thus, $I_j=0$ if and only if the corresponding marginals are deterministic. 

The overall degrees of indeterminism and signalling for the model may now be defined by
\begin{equation} \label{is}
I := \max \{I_1, I_2\},~~~S:= \max \{S_{1\rightarrow 2},S_{2\rightarrow 1}\} .
\end{equation}
These are easily checked to be equivalent to the less formal definitions in the introduction.  In particular, a model is deterministic and no-signalling if and only if $I=S=0$.

Note that the degrees of indeterminism and signalling are not completely independent of each other.  Indeed, any shift in a marginal probability value, $p$, due to signalling, must either keep the value in the same subinterval, $[0,I]$ or $[1-I,1]$ ($S\leq I$), or shift the value across the gap between the subintervals ($S\geq 1-2I$), leading to
\begin{equation} \label{imin}
I\geq \min \{ S, (1-S)/2 \} .
\end{equation}
This does not restrict the statement of the theorem in the following section, which refers to upper bounds for $I$ and $S$, rather than to the values of $I$ and $S$ for a particular model.

\section{A relaxed Bell inequality}

As noted in the introduction, the correlations generated by underlying deterministic no-signalling models must satisfy various Bell inequalities.  The degrees of determinism and no-signalling must therefore be relaxed to a certain extent, to model violations of these inequalities.  The extent of relaxation required may be quantified via corresponding `relaxed' Bell inequalities, such as the following (proved in the Appendix):
 
{\bf Theorem:} Let $X,X'$ and $Y,Y'$ denote possible measurement settings for a first and second observer, respectively, and label each measurement outcome by $1$ or $-1$. If $\langle XY\rangle$ denotes the average product of the measurement outcomes, for joint measurement settings $X$ and $Y$, then
\begin{equation} \label{bell}
 \langle XY\rangle + \langle XY'\rangle + \langle X'Y\rangle - \langle X'Y'\rangle   \leq B(I,S) 
\end{equation}
for any underlying model having values of indeterminism and signalling of at most $I$ and $S$, respectively, with tight upper bound
\begin{eqnarray} \nonumber
B(I,S) &=& 2+4 I~~~{\rm for}~~ S< 1-2I,\\ \label{bis}
&=& 4~~~~~~~~~~{\rm for}~~ S\geq 1-2I .
\end{eqnarray}

For deterministic no-signalling models, i.e., $I=S=0$, one has $B(0,0)=2$ and the theorem reduces to the well known Bell-CHSH inequality  \cite{chsh}.  More generally, if some value $B>2$ is observed for the lefthand side of Eq.~(\ref{bell}), the theorem places strong constraints on the minimum possible degrees of indeterminism and signalling that must be present in any corresponding model.  

In particular, consider a violation by an amount $V$ of the Bell-CHSH inequality, so that the lefthand side of Eq.~(\ref{bell}) equals $2+V$. Hence, any corresponding model must satisfy the complementarity relation 
\begin{equation} \label{comp2}
B(I,S)\geq 2+V .
\end{equation}
From Eq.~(\ref{bis}) this is equivalent to
\begin{equation} \label{v}
I\geq I_V := V/4   {\rm ~~and/or~~}  S\geq S_V := 1-V/2 .
\end{equation}

Since $V= 2\sqrt{2}-2$ for singlet state correlations \cite{root2}, it follows from Eq.~(\ref{v}) that any singlet state model must either assign no more than $1-I_V\approx 80\%$ certainty to some spin value, and/or predict a change of at least $S_V\approx 60\%$ in some underlying spin probability for one system in response to an operation performed on the other system.  Eqs.~(\ref{comp2}) and (\ref{v}) are thus weaker than the singlet state conjecture in Eq.~(\ref{comp}), but have the advantage of quantifying the complementarity of indeterminism and no-signalling for {\it any} violation and {\it any} physical system, not just for the singlet state.

A further consequence of the theorem is that the maximum possible violation of the Bell-CHSH inequality, $B(I,S)=4$, can be achieved only if 
\begin{equation} \label{sgap}
S\geq S_{\rm gap} := 1-2I ,
\end{equation}
i.e., only if Eq.~(\ref{comp}) holds.  Maximum violation for the particular case $I=1/2$ and $S=0$ (maximum indeterminism and no signalling) is achieved by the `PR-boxes' previously discussed in the literature \cite{rastall, pr}. Such boxes may be generalised to a continuum of $(I,S)$-boxes, each corresponding to a `nonlocal' resource saturating the upper bound $B(I,S)$.  Examples are given in Eqs.~(\ref{box1}) and (\ref{box2}) below.

The quantity $S_{\rm gap}$ in Eq.~(\ref{sgap}) corresponds to signalling of a magnitude sufficient to bridge the gap between the subintervals $[0,I]$ and $[1-I,1]$.  This is the minimum degree of signalling needed to be able to nonlocally flip some underlying marginal probability, $p$, to its complementary value, $1-p$. It is a surprising consequence of the theorem that signalling is {\it only} a useful resource, for the purposes of standard Bell inequality violation, when $S\geq S_{\rm gap}$. Note further that the conjecture in Eq.~(\ref{comp}) can be reformulated as the necessity of this `flipping' condition for any model of singlet state correlations.

\section{Information-theoretic complementarity}

The complementarity relations~(\ref{comp2}) and (\ref{v}) may also be formulated in terms of the number of random bits of local entropy and the number of nonlocal signalling bits of mutual information required to simulate a given Bell inequality violation.

First, for a model with degree of indeterminism $I$, it follows from Eqs.~(\ref{ilocal}) and (\ref{is}) that there is an underlying marginal probability distribution arbitrarily close to $(I,1-I)$.  Hence, there is some underlying outcome with Shannon entropy arbitrarily close to \begin{equation} \label{hi}
H(I):= - I\log_2 I - (1-I)\log_2 (1-I) .
\end{equation}
 
Second, for a model with degree of signalling $S$, it follows from Eqs.~(\ref{1to2}), (\ref{2to1}) and (\ref{is}) that, depending on one observer's choice between two measurement settings, the other observer will obtain an underlying marginal distribution of the form $(p, 1-p)$ or $(p+S,1-p-S)$.  If these two settings are selected between with corresponding equal prior probabilities of $1/2$, the Shannon mutual information which can in principle be communicated between the observers follows as
\[
 H(p+S/2) - (1/2) H(p) - (1/2) H(p+S) .
\]
This quantity is easily found to be minimised for the choice $p=(1-S)/2$.  Hence, it is 
is always possible, in principle, to communicate a Shannon mutual information of at least 
\begin{equation} \label{cs}
C(S):=1-H((1-S)/2) 
\end{equation}
bits per joint measurement.  Thus, $C(S)$ is the capacity of a symmetric binary channel based on the maximum available probability shift, $S$, due to signalling.

It follows immediately, from Eqs.~(\ref{v}), (\ref{hi}) and (\ref{cs}), that any underlying model or simulation of a violation $V$ of the Bell-CHSH inequality requires the generation of at least either 
\begin{equation} \label{hv}
H_V := H(I_V)=  H(V/4) 
\end{equation}
random bits of local entropy, and/or communication of
\begin{equation} \label{cv}
C_V := C(S_V):= 1-H(V/4) = 1 - H_V
\end{equation}
nonlocal signalling bits of mutual information.  For $V= 2\sqrt{2}-2$ one has $H_V\approx 0.736$ bits and $C_V\approx 0.264$ bits.  

Finally, note that the conjecture in Eq.~(\ref{comp}) is equivalent to the information complementarity relation
\begin{equation} \label{conjinf} 
C(S) + H(I) \geq 1
\end{equation}
for models of singlet state correlations.  A similar relation has very recently been independently conjectured \cite{gar}.

\section{Support for conjecture}

As noted in the introduction, the singlet state conjecture in Eq.~(\ref{comp}), and hence the information complementarity relation in Eq.~(\ref{conjinf}), is certainly valid for $S=0$, as a consequence of the result by Branciard et al. \cite{branc}.  Further, it may be shown that either of Eqs.~(\ref{comp}) and (\ref{conjinf}) are {\it sufficient} to ensure that all spin correlations can be modelled for the singlet state.
The latter follows by exhibiting explicit models for which $S+2I=1=H(I)+C(S)$.  

In particular, consider the models corresponding to mixtures of the (deterministic) Toner-Bacon model \cite{bt} with the standard (no-signalling) Hilbert space model:
\[ p_{XY}(a,b|\lambda):=w\, p^{(BT)}_{XY}(a,b|\lambda) + (1-w) (1-ab\, x\cdot y)/4 \]
for $0\leq w\leq 1$ (where measurement setting $X$ is associated with spin direction $x$, and $\lambda$ denotes two unit 3-vectors \cite{bt}).  Hence, the marginals are restricted to the values $(1\pm w)/2$, yielding $S=w=1-2I$, and $H(I)=H((1-w)/2)=1-C(S)$, as required.  

It is also straightforward to show that the conditions in Eqs.~(\ref{comp}) and (\ref{conjinf}) are in fact {\it necessary} for any singlet state model with $S\geq 1/3$.  For this range of $S$ one has $1-S\leq 2S$.  Hence,  $I\geq (1-S)/2$ from Eq.~(\ref{imin}), which is equivalent to Eq.~(\ref{comp}), and further implies Eq.~(\ref{conjinf}) via the monotonicity of $H(I)$. 

Thus, to completely prove the conjecture, it remains to be shown that there is no model of singlet state correlations for the case $0<S<1/3$ and $S+2I<1$.  

\section{Discussion}

Complementarity relations for the contributions of indeterminism and signalling to any violation of the Bell-CHSH inequality have been obtained in Secs.~III and IV, via the `relaxed' Bell inequality of the theorem.  Stronger complementarity relations, (\ref{comp}) and (\ref{conjinf}), for the contributions required in any model of the singlet state, have been conjectured, and partially proved in Sec.~V.

The results suggest that a common practice of referring to Bell inequality violation as evidence of `quantum nonlocality' (and to `nonlocal' boxes) is conceptually unclear.  For example, in the no-signalling case, $S=0$, it is not apparent in what sense violations arising from an underlying indeterminism can be considered `nonlocal'.  Further, even for $S>0$, there is no clear distinction between `quantum nonlocality' and `classical nonlocality' (where the latter concept includes, for example, action-at-a-distance models such as Newtonian gravity).  This supports critiques by Mermin \cite{mermin}, who concluded that use of the term `quantum nonlocality' is no more than `fashion at a distance'; and by Kent \cite{kent}, who  lamented it as a `confusingly oxymoronic phrase' that conflicts with the notion of quantum locality in field theory (see also comments by Zukowski \cite{zuk} and Griffiths \cite{griffiths}).   Further, the alternative term `Bell nonlocality'  is at best tautological, throwing no light on why Bell inequality violations {\it per se} should be characterised as `nonlocal' (particularly for $S=0$).  To capture the notion that the statistics underlying such violations cannot be reduced to independent contributions from each measurement region, it would perhaps be clearer to refer to `nonseparable' or `irreducible' correlations. 

There are a number of possible generalisations of the above results that might be fruitfully explored, in addition to relaxing other known Bell inequalities and to proving the conjecture in full.  

First, one could consider {\it asymmetric} restrictions on the degrees of local indeterminism and one-way signalling in Eqs.~(\ref{1to2})-(\ref{ilocal}).  For example, if the first observer's measurements are in the past of the second observer's, it is natural to take $S_{1\rightarrow 2}>0$ and $S_{2\rightarrow 1}=0$. 

Second, it may be possible to sharpen the information-theoretic bounds in Eqs.~(\ref{hv}) and (\ref{cv}) via appropriate redefinitions of $I$ and $S$.  For example, one could redefine $I$ as the minimum entropy  of any underlying marginal (or joint) distribution, and $S$ as the maximum underlying mutual information obtainable via one observer varying the choice of measurement settings.  

Third, it would be of interest to further investigate connections with communication models of quantum correlations, which are typically deterministic but signalling  \cite{bt,tess,regev,paw}.  For example, the mixed model considered in Sec.~V shows how such models may be `relaxed', allowing a trade-off between the local number of bits generated and the non-local number of bits communicated.  Further, it is remarkable that the number of locally generated randoms bits, $H_V=0.736$, required for simulating the maximum quantum violation of the Bell-CHSH inequality via a no-signalling indeterministic model, as per Sec.~IV, is in fact equal to the number of nonlocal signalling bits required in the signalling deterministic model given recently by  Pawlowski et al. \cite{paw}.

Finally, while the above results rely on the assumption that measurement settings may be chosen freely, one could quantify a corresponding degree of relaxation, by some quantity $M$, and generalise the above theorem to a suitable tight bound $B(I,S,M)$.  For example, from Eq.~(\ref{underlying}) one might define $M:=\sup_{X,X',Y,Y'} \int d\lambda \, |\rho_{XY}(\lambda) - \rho_{X'Y'}(\lambda)|$  (the bound $B(I,S,M) = 4-(2-3M)(1-2I)$ for $S\leq 1-2I$ and $M\leq 2/3$, and $=4$ otherwise, is conjectured).

{\it Acknowledgment:} I thank Prof. Gisin for helpful comments regarding Refs.~\cite{bt,branc}, and the referee for suggesting useful clarifications.
 
\appendix

\section{Obtaining the theorem}

The relaxed Bell inequality in Sec.~III is proved here.  First, for outcomes labelled by $\pm 1$, the joint measurement outcomes may be ordered as $(+,+)$, $(+,-)$, $(-,+)$, and $(-,-)$.  The corresponding probability distribution $p_{XY}(a,b|\lambda)$ may then be written as
$(c, m-c, n-c, 1+c-m-n)$, where $m$ and $n$ denote the corresponding marginal probabilities $p^{(1)}_{XY}(+|\lambda)$ and $p^{(2)}_{XY}(+|\lambda)$.  
Thus, if $\langle XY\rangle_\lambda$ denotes the average product of the measurement outcomes for a fixed value of $\lambda$, then $\langle XY\rangle_\lambda = 1 + 4c-2(m+n)$.  Positivity of joint probabilities requires
$\max\{ 0, m+n-1\} \leq c \leq \min\{m,n\}$.  Hence, noting $2\,\max(x,y) = x+y+|x-y|$,
\[     2|m+n-1|-1\leq \langle XY\rangle_\lambda  \leq 1-2|m-n|, \]
where the upper and lower bounds are attainable via suitable choices of $c$.  

Writing $p_1\equiv p_{XY}$, $p_2\equiv p_{X'Y}$, $p_3\equiv p_{XY'}$, $p_4\equiv p_{X'Y'}$, and 
$E_\lambda := \langle XY\rangle_\lambda + \langle XY'\rangle_\lambda + \langle X'Y\rangle_\lambda - \langle X'Y'\rangle_\lambda$,
it immediately follows that
\begin{equation} \label{elam}
E_\lambda\leq 4-2J ,
\end{equation} 
where
\[
 J:= |m_1-n_1| + |m_2-n_2| + |m_3-n_3| + |m_4 + n_4-1| ,
\]
and the upper bound is attainable for suitable choices of $c_1$, $c_2$, $c_3$ and $c_4$.  
The indeterminism and signalling constraints of the theorem imply that
$m_j,n_j \in [0,I]\cup [1-I,1]$  and $|m_1-m_2|, |m_3-m_4|, |n_1-n_3|, |n_2-n_4| \leq S$.  To maximise $E_\lambda$ it is therefore necessary to minimise $J$ under these constraints.

Suppose first that $S< 1-2I$.  Hence, the pair $\{ m_1, m_2\}$ is a subset of one of the two subintervals $[0,I]$ and $[1-I,1]$ (since signalling cannot bridge the gap), as are the pairs $\{m_3,m_4\}$, $\{n_1,n_3\}$ and $\{n_2,n_4\}$.  Now, if $m_4$ and $n_4$ share the same subinterval, then $|m_4+n_4-1|\geq 1-2I$, with equality for $m_4=n_4=I$ (or $1-I$). Hence $J\geq 1-2I$, with equality for $m_j=n_j=I$ (or $1-I$) for each $j$.  Otherwise, $m_4$ and $n_4$, and hence their respective paired partners, $m_3$ and $n_2$, lie in different subintervals.  

In this latter case, note first that if $m_1$ and $n_1$ also lie in different subintervals, then $|m_1-n_1|\geq 1-2I$, with equality for $m_1=1-n_1=I$ (or $1-I$).  Hence, $J\geq 1-2I$, with equality for $m_1=m_2=n_2=n_4=1-m_3=1-m_4=1-n_1=1-n_3=I$ (or $1-I$).    Conversely, if $m_1$ and $n_1$ lie in the {\it same} subinterval, then so must their respective paired partners $m_2$ and $n_3$, i.e., the set $\{ m_1,m_2,n_1,n_3\}$ lies in the same subinterval.  But from the last sentence of the above paragraph, this can only  be the same subinterval as {\it one} of the two pairs $\{m_3,m_4\}$ and $\{n_2,n_4\}$.  Hence, either $m_2$ and $n_2$, or $m_3$ and $n_3$, lie in different subintervals, implying that $|m_2-n_2|+|m_3-n_3| \geq 1-2I$, with equality if $m_2=n_2$ and $m_3=1-n_3=I$ (or $1-I$), or if $m_3=n_3$ and $m_2=1-n_2=I$ (or $1-I$).  Hence, $J\geq  1-2I$, with equality for $m_1=m_2=1-m_3=1-m_4=n_j=I$ (or $1-I$) for all $j$, or for $m_j=n_1=n_3=1-n_2=1-n_4=I$ (or $1-I)$ for all $j$.  

Thus, for $S<1-2I$, one has the tight bound $J\geq1-2I$ in all cases, implying via Eq.~(\ref{elam}) the tight bound $E_\lambda\leq 2+4I$. Finally, if $S\geq 1-2I$, one trivially has $J\geq 0$ (with equality for $m_j=n_j=m_4=1-n_4=I$ (or $1-I$) for $j=1,2,3$), yielding the tight bound $E_\lambda \leq 4$.  
Averaging these upper bounds for $E_\lambda$ over $\lambda$ gives the theorem as desired.

Note that sets of joint probability distributions which saturate the upper bound $B(I,S)$ of the theorem follow from the examples of equality given above.  One has, for example, the $(I,0)$-box
\begin{equation} \label{box1}
p_1\equiv p_2\equiv p_3 \equiv (I,0,0,1-I), ~p_4\equiv (0,I,I,1-2I) 
\end{equation}
with $B(I,0)=4I$, and the $(I,1-2I)$-box
\begin{equation} \label{box2}
p_1\equiv p_2\equiv p_3 \equiv (I,0,0,1-I), ~p_4\equiv (0,I,1-I,0) 
\end{equation}
with $B(I,1-2I)=4$.  Each of these reduces to the standard PR-box when $I=1/2$ \cite{rastall}.


\newpage



\end{document}